\documentclass[conference]{IEEEtran}

\usepackage{cite}
\usepackage{amsmath,amssymb,amsfonts}
\usepackage{algorithmic}
\usepackage{graphicx}
\usepackage{textcomp}
\usepackage{xcolor}
\usepackage{algorithm}
\usepackage{booktabs}
\usepackage{caption}

\usepackage{tikz}
\usetikzlibrary{positioning, fit, calc, shapes, arrows.meta, backgrounds, shadows}
\usepackage{hyperref} 

\hypersetup{
    colorlinks=true,      
    linkcolor=blue,       
    citecolor=blue,       
    urlcolor=blue,        
    pdftitle={My Paper},  
    pdfauthor={My Name}   
}

\begin{document}

\title{Sparse Broad Learning System via 
Sequential Threshold Least-Squares for 
Nonlinear System Identification under Noise}

\author{\IEEEauthorblockN{Zijing Li}
\IEEEauthorblockA{\textit{School of Astronautics} \\
\textit{Harbin Institute of Technology}\\
Harbin, China \\
24S104221@stu.hit.edu.cn}
\and
\IEEEauthorblockN{Second Author}
\IEEEauthorblockA{\textit{School of Astronautics} \\
\textit{Harbin Institute of Technology}\\
Harbin, China \\
email@address.com}
}

\maketitle

\begin{abstract}
The Broad Learning System (BLS) has gained 
significant attention for its computational 
efficiency and less network parameters 
compared to deep learning structures. 
However, the standard BLS relies on the 
pseudoinverse solution, which minimizes the mean square error with
$L_2$-norm but lacks robustness against 
sensor noise and outliers common in industrial 
environments. 
To address this limitation, this paper proposes 
a novel \textit{Sparse Broad Learning System} 
(S-BLS) framework. 
Instead of the traditional ridge regression, 
we incorporate the Sequential Threshold 
Least-Squares (STLS) algorithm—originally 
utilized in the sparse identification of 
nonlinear dynamics (SINDy)—into the output 
weight learning process of BLS. 
By iteratively thresholding small coefficients, 
the proposed method promotes sparsity in the 
output weights, effectively filtering out 
noise components while maintaining modeling 
accuracy. 
This approach falls under the category of 
sparse regression and is particularly 
suitable for noisy environments. 
Experimental results on a numerical nonlinear 
system and a noisy Continuous Stirred Tank 
Reactor (CSTR) benchmark demonstrate that 
the proposed method is effective and achieves 
superior robustness compared to standard BLS.
\end{abstract}

\section{Introduction}

Nonlinear system identification plays a pivotal role in modern industrial applications, such as chemical process control and robotics, where capturing complex dynamics from data is essential. 
While Deep Learning (DL) approaches, including Recurrent Neural Networks (RNNs) and Long Short-Term Memory (LSTM) networks, have demonstrated powerful approximation capabilities, they are often plagued by computationally intensive training processes, massive data requirements, and the need for extensive hyperparameter tuning. 
To address these limitations and offer a computationally efficient alternative, the Broad Learning System (BLS) was proposed by Chen and Liu \cite{chen2017bls}. 
Unlike deep architectures that stack layers vertically, BLS expands nodes horizontally and determines output weights analytically using the Moore-Penrose pseudoinverse. 
This unique flat structure ensures extremely fast training speeds, making BLS highly suitable for real-time industrial modeling.

To enhance the performance of standard BLS, extensive research has been conducted regarding structural evolution, robust learning, and regularization.

From a structural perspective, various architectures have been developed to adapt BLS to complex tasks. 
Inspired by the depth concept, Stacked BLS \cite{StackedBLS2021} was introduced to enhance feature abstraction while maintaining the system's learning ability. 
Similarly, to handle uncertainty and dynamic variations, Fuzzy BLS (FBLS) \cite{FBLS1020} and Recurrent BLS (RBLS) \cite{RBLS2018} have been proposed. 
However, these structural expansions often result in a substantial increase in the number of network nodes, leading to model redundancy and higher computational costs during the inference phase.

Regarding optimization objectives, standard BLS relies on the $L_2$-norm (Mean Square Error), which is sensitive to non-Gaussian noise and outliers commonly found in industrial environments. 
To mitigate this, robust variants such as the Maximum Correntropy Criterion (MCC-BLS) \cite{MCCBLS2020} and Weighted BLS (WBLS) \cite{weightedBLS2019} have been introduced. 
WBLS improves robustness by assigning lower weights to outliers. 
\textbf{Nevertheless, these methods face two critical limitations:} 
First, while they reduce the impact of noise, they typically retain the full network topology without eliminating redundant nodes (i.e., lack of sparsity). 
Second, the introduction of complex non-convex loss functions or iterative weighting often precludes the use of the pseudoinverse closed-form solution, thereby sacrificing the core speed advantage of BLS.

To induce sparsity, regularization techniques have been applied. 
Initially, sparse models based on the $L_1$-norm (Lasso), such as the Compact BLS \cite{compactbls2023}, were developed. 
While $L_1$ regularization can suppress insignificant nodes, it acts as a "soft" thresholding operator, which may not sufficiently prune small coefficients caused by noise.
More recently, direct $L_0$-norm optimization has been explored. 
Chu et al. proposed a Controllable Sparse BLS (CSBLS) \cite{CSBLS2023}, which enforces sparsity via a cardinality constraint ($\|W\|_0 \le Q$) and solves it using Normalized Iterative Hard Thresholding (NIHT). 
Although CSBLS effectively controls model size, it fundamentally relies on a first-order gradient descent approach, which can be slow to converge compared to analytical methods. 
Furthermore, CSBLS requires the number of active nodes $Q$ to be fixed \textit{a priori}. 
In system identification, the true order of the system is often unknown, making it difficult to pre-determine the optimal $Q$ without extensive trial and error.

In the field of data-driven discovery of dynamics, the Sparse Identification of Nonlinear Dynamics (SINDy) \cite{sindy2016} has demonstrated that finding a sparse representation can effectively filter out noise. 
The core solver of SINDy is the Sequential Threshold Least-Squares (STLS) algorithm. 
Unlike soft regularization or gradient-based pruning, STLS acts as a "hard" thresholding operator combined with least-squares projection. 
It rapidly converges to a sparse solution by iteratively zeroing out coefficients smaller than a noise threshold $\lambda$. 
This mechanism allows the model to focus solely on the underlying dynamics while discarding noise-induced connections.

Inspired by the efficiency of STLS, we propose a robust \textbf{Sparse Broad Learning System (S-BLS)}. 
Instead of imposing a fixed node constraint like CSBLS or employing slow iterative solvers, we integrate the STLS algorithm directly into the BLS weight optimization process. 
This approach differs from existing methods in that it adaptively determines the active feature nodes based on the noise level, allowing S-BLS to simultaneously achieve structural sparsity and noise robustness while maintaining the computational efficiency characteristic of the original BLS.

The main contributions of this paper are summarized as follows:
\begin{enumerate}
    \item We propose a unified framework that integrates the Sequential Thresholded Least-Squares (STLS) algorithm into BLS. This effectively replaces the dense $L_2$ pseudoinverse solution with a sparse, noise-resilient solution. Unlike CSBLS \cite{CSBLS2023}, our method does not require a pre-set number of nodes and avoids slow gradient-based optimization.
    
    \item The proposed method exhibits superior robustness and interpretability. The STLS mechanism acts as a "hard" filter, automatically pruning redundant feature nodes corrupted by noise, which significantly reduces the model complexity.
    
    \item We validate the proposed S-BLS on both numerical nonlinear identification tasks and a highly nonlinear Continuous Stirred Tank Reactor (CSTR) process. The experimental results demonstrate that S-BLS outperforms standard BLS and its robust variants (e.g., WBLS) in terms of both prediction accuracy and noise immunity.
\end{enumerate}
\section{Preliminaries}

\subsection{Broad Learning System (BLS)}
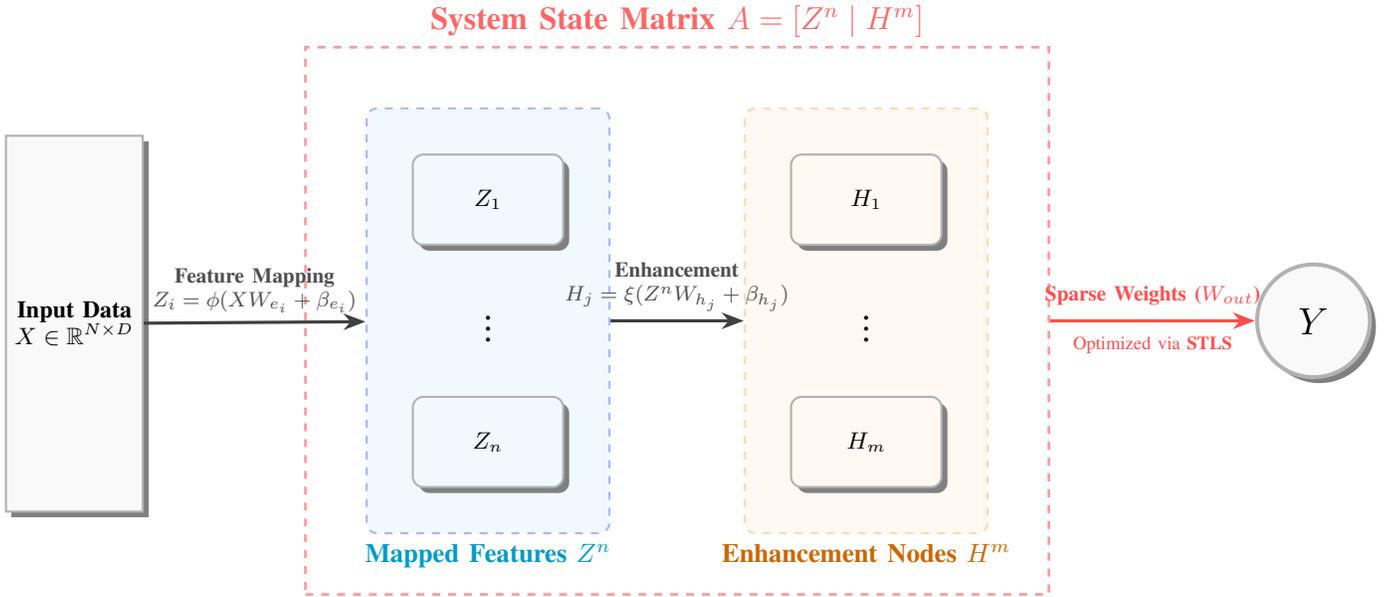
\begin{figure*}[htbp] 
\centering

\begin{tikzpicture}[
    node distance=1.5cm and 3.5cm,
    every node/.style={font=\small},
    group/.style={draw=blue!30, rounded corners, inner sep=0.6cm, fill=blue!5, dashed, thick},
    input/.style={draw=gray!60, rectangle, thick, minimum width=1.8cm, minimum height=5cm, align=center, fill=gray!5, drop shadow},
    block/.style={draw=gray!60, rectangle, rounded corners, minimum width=2cm, minimum height=1.2cm, align=center, fill=white, drop shadow, thick},
    arrow/.style={-Stealth, line width=1.2pt, color=darkgray},
    label text/.style={font=\footnotesize\bfseries, color=black!70, align=center, midway, above}
]

\node[input] (X) {\textbf{Input Data}\\$X \in \mathbb{R}^{N \times D}$};

\node[block, fill=cyan!5, right=5.5cm of X.north, anchor=north, yshift=-0.25cm] (Z1) {$Z_1$};
\node[below=0.6cm of Z1] (Zdots) {\Large $\vdots$};
\node[block, fill=cyan!5, below=0.6cm of Zdots] (Zn) {$Z_n$};

\begin{scope}[on background layer]
    \node[group, fit=(Z1) (Zn), fill=cyan!5, label={[cyan!80!black, font=\bfseries]below:Mapped Features $Z^n$}] (Z_layer) {};
\end{scope}

\node[block, fill=orange!5, right=3cm of Z1] (H1) {$H_1$};
\node[below=0.6cm of H1] (Hdots) {\Large $\vdots$};
\node[block, fill=orange!5, below=0.6cm of Hdots] (Hm) {$H_m$};

\begin{scope}[on background layer]
    \node[group, fit=(H1) (Hm), fill=orange!5, draw=orange!30, label={[orange!80!black, font=\bfseries]below:Enhancement Nodes $H^m$}] (H_layer) {};
\end{scope}

\node[draw=red!40, dashed, fit=(Z_layer) (H_layer), inner sep=0.8cm, line width=1pt, 
      label={[red!60, font=\large\bfseries]above:System State Matrix $A = [Z^n \mid H^m]$}] (A_matrix) {};

\node[right=3.5cm of A_matrix.east, anchor=center] (Y) [draw=gray!60, circle, line width=1.5pt, minimum size=1.5cm, fill=gray!5, drop shadow] {\Large $Y$};

\draw[arrow] (X) -- (Z_layer) node[label text] {Feature Mapping\\$Z_i = \phi(XW_{e_i} + \beta_{e_i})$};
\draw[arrow] (Z_layer) -- (H_layer) node[label text] {Enhancement\\$H_j = \xi(Z^n W_{h_j} + \beta_{h_j})$};
\draw[arrow, color=red!70] (A_matrix.east) -- (Y) 
    node[midway, above, font=\footnotesize\bfseries, color=red!70, align=center, yshift=2pt] {Sparse Weights ($W_{out}$)}
    node[midway, below, font=\scriptsize, color=red!70, align=center, yshift=-2pt] {Optimized via \textbf{STLS}};

\end{tikzpicture}
\caption{The architecture of the proposed Sparse Broad Learning System (S-BLS). The input $X$ is mapped to feature nodes $Z^n$, which are cascaded to enhancement nodes $H^m$. Both groups are concatenated to form the system matrix $A$. The output weights $W_{out}$ are solved sparsely using the Sequential Thresholded Least-Squares (STLS) algorithm.}
\label{fig:bls_structure}
\end{figure*}

The Broad Learning System (BLS) is built upon the concept of the Random Vector Functional Link Neural Network (RVFLNN) \cite{RVFLNN1994}. 
It eliminates the need for time-consuming backpropagation by fixing the input weights and optimizing only the output weights. 
Theoretically, BLS ensures universal approximation capability, meaning it can approximate any continuous function to arbitrary accuracy given a sufficient number of nodes.

Given input data $X \in \mathbb{R}^{N \times D}$, where $N$ is the number of samples and $D$ is the input dimension, the mapped feature nodes for the $i$-th group, denoted as $Z_i$, are generated as:
\begin{equation}
Z_i = \phi(X W_{e_i} + \beta_{e_i}), \quad i=1,\dots,n
\end{equation}
Here, $n$ denotes the total number of feature mapping groups. Assuming each group contains $k$ nodes, the random weights and biases are defined as $W_{e_i} \in \mathbb{R}^{D \times k}$ and $\beta_{e_i} \in \mathbb{R}^{k}$, respectively. Consequently, the $i$-th mapped feature block has the dimension $Z_i \in \mathbb{R}^{N \times k}$.
The complete mapped feature matrix is defined by concatenating all groups:
\begin{equation}
Z^n = [Z_1, Z_2, \dots, Z_n] \in \mathbb{R}^{N \times nk}
\end{equation}

These mapped features are then fed into enhancement nodes. The $j$-th group of enhancement nodes $H_j$ is calculated as:
\begin{equation}
H_j = \xi(Z^n W_{h_j} + \beta_{h_j}), \quad j=1,\dots,m
\end{equation}
where $\xi(\cdot)$ is the activation function (typically $\tanh$), and $m$ represents the number of enhancement groups. 
If each enhancement group consists of $q$ nodes, the connecting weights are $W_{h_j} \in \mathbb{R}^{nk \times q}$ and biases are $\beta_{h_j} \in \mathbb{R}^{q}$. Note that the input dimension for the enhancement layer corresponds to the total number of feature nodes, $nk$.
The enhancement nodes matrix is formulated as:
\begin{equation}
H^m = [H_1, H_2, \dots, H_m] \in \mathbb{R}^{N \times mq}
\end{equation}

The final system state matrix $A$, which serves as the regressor, is constructed by concatenating both mapped features and enhancement nodes:
\begin{equation}
A = [Z^n \mid H^m] \in \mathbb{R}^{N \times L}
\end{equation}
where $L = nk + mq$ denotes the total number of nodes in the hidden layer.

The relationship between the input matrix $A$ and the target output $Y \in \mathbb{R}^{N \times C}$ (where $C$ is the output dimension) is modeled linearly as $Y = A W_{out}$. 
The objective of standard BLS is to minimize the structural risk using Ridge Regression ($L_2$-regularization):
\begin{equation}
\min_{W_{out}} \|Y - A W_{out}\|_2^2 + \lambda \|W_{out}\|_2^2
\end{equation}
where $W_{out} \in \mathbb{R}^{L \times C}$ is the output weight matrix and $\lambda$ is the regularization parameter. 
The optimal solution is analytically given by the ridge inverse:
\begin{equation}
W_{out} = (A^T A + \lambda I)^{-1} A^T Y
\label{eq7}
\end{equation}

\textbf{Remark:} Although Equation \eqref{eq7} provides a rapid closed-form solution, it fundamentally relies on the $L_2$-norm. This results in a \textit{dense} weight matrix $W_{out}$, where almost all elements are non-zero. In noisy industrial environments, this dense structure forces the model to fit not only the system dynamics but also the noise, leading to poor generalization and a lack of interpretability.

\subsection{Sparse Optimization Problem}
To address the overfitting issue caused by the dense solution of standard BLS, a sparse representation is desirable. 
In the context of system identification, we seek a parsimonious model where only the most relevant nodes contribute to the output. 
Mathematically, this is formulated as an optimization problem with an $L_0$-norm penalty:
\begin{equation}
\min_{W} \|Y - A W\|_2^2 + \alpha \|W\|_0
\end{equation}
where $W \in \mathbb{R}^{L \times C}$ is the sparse weight matrix we aim to solve, $\|W\|_0$ counts the number of non-zero elements in $W$, and $\alpha$ is a parameter controlling the sparsity level.

However, directly solving Equation (8) is known to be an NP-hard combinatorial problem, making it intractable for large-scale systems. 
While relaxations such as $L_1$-regularization (Lasso) exist, they typically act as "soft" thresholding operators, which may shrink coefficients without eliminating them completely, and often require slow iterative solvers (e.g., ADMM or coordinate descent). 
Therefore, an efficient algorithm capable of solving the sparse identification problem within the fast computing framework of BLS is highly demanded.
\section{Proposed Method: S-BLS}
\label{sec:method}

In this section, we present the Sparse Broad Learning System (S-BLS). 
Existing robust variants of BLS primarily rely on weighting samples or complex non-convex loss functions, which often compromise the computational efficiency of the original framework. 
To address this, we propose a structural sparsity approach. 
By integrating a sequential thresholding strategy into the BLS learning paradigm, S-BLS automatically identifies and prunes redundant feature nodes corrupted by noise while maintaining a rapid closed-form solution.
The S-BLS structure is displayed in Fig. \ref{fig:bls_structure}.

\subsection{Model Formulation}
Consider the augmented feature matrix $A = [Z^n | H^m] \in \mathbb{R}^{N \times L}$ generated by the BLS structure, where $L$ is the total number of nodes ($L=nk+mq$).
In standard BLS, the output weight matrix $W \in \mathbb{R}^{L \times C}$ is obtained by minimizing the $L_2$-regularized error:
\begin{equation}
\min_{W} \mathcal{J}_{BLS}(W) = \frac{1}{2} \|Y - AW\|_F^2 + \frac{\lambda_{ridge}}{2} \|W\|_F^2
\end{equation}
where $\|\cdot\|_F$ denotes the Frobenius norm and $\lambda_{ridge}$ is the regularization parameter.
While this formulation yields a closed-form solution $W = (A^T A + \lambda_{ridge} I)^{-1} A^T Y$, the resulting weight matrix is \textit{dense}. 
In scenarios with heavy noise, the model tends to assign non-zero weights to all $L$ nodes to fit the noise, leading to overfitting.

To enhance robustness and interpretability, we introduce an $L_0$-norm constraint to enforce sparsity. The objective function of S-BLS is formulated as:
\begin{equation}
\min_{W} \mathcal{J}_{S-BLS}(W) = \frac{1}{2} \|Y - AW\|_F^2 + \alpha \|W\|_0
\label{eq10}
\end{equation}
where $\|W\|_0$ counts the number of non-zero elements in $W$, and $\alpha$ is the sparsity penalty parameter controlling the trade-off between error and model complexity.
Since optimizing the $L_0$-norm is NP-hard, standard gradient-based methods cannot be applied directly.

\subsection{Sequential Thresholding Optimization}
To solve the optimization problem in \eqref{eq10} efficiently, we adopt a \textbf{Sequential Thresholded Least-Squares (STLS)} strategy. 
Unlike soft-thresholding methods—such as the Iterative Shrinkage-Thresholding Algorithm (ISTA) or Lasso—that shrink coefficients continuously, STLS employs a hard thresholding operator combined with iterative projections.

Let $W^{(t)} \in \mathbb{R}^{L \times C}$ denote the weight matrix at iteration $t$. The optimization procedure consists of two alternating steps:

\subsubsection*{Step 1: Hard Thresholding (Pruning)}
We define a hard thresholding operator $\mathcal{T}_{\lambda}(\cdot)$. 
For each element $w_{ij}$ in the weight matrix, small coefficients dominated by noise are truncated to zero:
\begin{equation}
w_{ij}^{(t)} = \mathcal{T}_{\lambda}(w_{ij}^{(t-1)}) = 
\begin{cases} 
0, & \text{if } |w_{ij}^{(t-1)}| < \lambda \\
w_{ij}^{(t-1)}, & \text{otherwise}
\end{cases}
\end{equation}
where $\lambda$ is the explicit truncation threshold derived from the sparsity parameter $\alpha$.
This step identifies the \textit{active set} of nodes, denoted as $\mathcal{S}^{(t)}$, which contains the indices of the retained significant features for each output dimension.

\textbf{Physical Interpretation:} In the context of BLS, this thresholding step acts as a \textit{structural pruner}. 
Since $A$ is composed of mapped features and enhancement nodes, setting a row in $W$ to zero is equivalent to physically removing the corresponding node connection from the network. 
This allows S-BLS to automatically discard invalid feature nodes that capture only noise.

\subsubsection*{Step 2: Projection on Active Set}
Once the insignificant nodes are pruned, we update the weights of the remaining nodes to minimize the reconstruction error. 
This is achieved by solving the least-squares problem restricted to the active set $\mathcal{S}^{(t)}$:
\begin{equation}
W_{\mathcal{S}}^{(t)} = \arg\min_{W} \|Y - A_{\mathcal{S}} W\|_F^2 = (A_{\mathcal{S}}^T A_{\mathcal{S}})^{-1} A_{\mathcal{S}}^T Y
\end{equation}
where $A_{\mathcal{S}} \in \mathbb{R}^{N \times |\mathcal{S}|}$ denotes the sub-matrix of $A$ formed by extracting only the columns corresponding to the active indices in $\mathcal{S}^{(t)}$, and $|\mathcal{S}|$ represents the number of active nodes.
Elements not in the active set remain zero.

\subsubsection*{Convergence and Stopping Criterion}
Instead of relying on a small perturbation threshold $\epsilon$ for convergence, which may lead to prolonged computation in oscillatory cases, we employ a **fixed-iteration** strategy. 
Since the STLS algorithm utilizes the Moore-Penrose pseudoinverse for projection, it finds the optimal solution within the current subspace in a single step. 
Empirical studies observe that the active support set $\mathcal{S}$ typically stabilizes extremely quickly. 
Therefore, we set a maximum iteration count $T$ (e.g., $T=10$) as the stopping criterion. 
This ensures strictly bounded computational time, which is critical for industrial applications.

The complete training procedure is summarized in Algorithm \ref{alg:SBLS}.

\begin{algorithm}[htbp]
\caption{Training Algorithm for S-BLS}
\label{alg:SBLS}
\begin{algorithmic}[1]
\REQUIRE Input $X \in \mathbb{R}^{N \times D}$, Output $Y \in \mathbb{R}^{N \times C}$; 
\REQUIRE BLS Hyperparameters ($n, m$); Sparsity Threshold $\lambda$; Max Iterations $T$.
\ENSURE Sparse Output Weights $W \in \mathbb{R}^{L \times C}$.

\STATE \textbf{1. Initialization:}
\STATE Randomly generate weights $W_{e_i}, \beta_{e_i}$ and $W_{h_j}, \beta_{h_j}$.
\STATE Compute feature nodes $Z^n$ and enhancement nodes $H^m$.
\STATE Construct system matrix $A = [Z^n \mid H^m] \in \mathbb{R}^{N \times L}$.
\STATE Compute initial dense solution: $W^{(0)} = A^{\dagger} Y$.

\STATE \textbf{2. Sequential Thresholding Loop:}
\FOR{$t = 1$ to $T$}
    \STATE \textit{// Step A: Hard Thresholding}
    \STATE Identify small coefficients: $small\_idx \leftarrow \{ (i,j) : |W_{ij}^{(t-1)}| < \lambda \}$.
    \STATE Prune weights: $W^{(t)} \leftarrow W^{(t-1)}$.
    \STATE $W^{(t)}[small\_idx] \leftarrow 0$.
    
    \STATE \textit{// Step B: Update Active Set}
    \FOR{each output dimension $d = 1$ to $C$}
        \STATE Identify active indices: $\mathcal{S}_d \leftarrow \{ i : W_{i,d}^{(t)} \neq 0 \}$.
        \STATE Construct sub-matrix $A_{\mathcal{S}_d} \in \mathbb{R}^{N \times |\mathcal{S}_d|}$.
        \STATE Update non-zero weights via Least Squares:
        \STATE $W^{(t)}_{\mathcal{S}_d, d} = (A_{\mathcal{S}_d}^T A_{\mathcal{S}_d})^{-1} A_{\mathcal{S}_d}^T Y_{:,d}$.
    \ENDFOR
\ENDFOR

\RETURN Sparse weights $W^{(T)}$.
\end{algorithmic}
\end{algorithm}

\subsection{Complexity Analysis}
We compare the computational complexity of the proposed S-BLS with standard BLS and Lasso-BLS. Let $N$ be the number of samples and $L$ be the total number of nodes ($L = nk + mq$). Assume $N > L$.

\begin{enumerate}
    \item \textbf{Standard BLS:} Requires one pseudoinverse calculation. The complexity involves calculating the correlation matrix $A^T A$ and its inversion. Thus, the complexity scales as $\mathcal{O}(NL^2 + L^3)$. While computationally fast, standard BLS lacks sparsity, leading to dense matrices.
    
    \item \textbf{Lasso-BLS ($L_1$):} Solves the sparse optimization problem using iterative algorithms such as the Alternating Direction Method of Multipliers (ADMM) or Iterative Shrinkage-Thresholding Algorithm (ISTA). These methods typically require a large number of iterations ($K_{lasso}$) to converge. Each iteration involves matrix-vector multiplications with complexity $\mathcal{O}(NL)$. Therefore, the total complexity is roughly $\mathcal{O}(K_{lasso} \cdot NL)$. Since $K_{lasso}$ is usually large (e.g., hundreds of iterations), this approach is computationally expensive for large-scale datasets.
    
    \item \textbf{Proposed S-BLS:} The STLS algorithm is also iterative but employs a fixed, small number of iterations $T$ (e.g., $T=10$). The initialization step requires $\mathcal{O}(NL^2 + L^3)$. However, in subsequent iterations, the number of active nodes $L_{active}$ decreases significantly ($L_{active} \ll L$). The complexity of the update step in Eq. (13) reduces to $\mathcal{O}(N L_{active}^2 + L_{active}^3)$. Given that $T$ is small and $L_{active}$ shrinks rapidly, the additional computational cost over standard BLS is marginal, and the total complexity remains far lower than that of Lasso-BLS.
\end{enumerate}


\section{Experiments and Analysis}
\label{sec:experiments}

In this section, we evaluate the proposed S-BLS on two benchmark tasks: a numerical nonlinear system identification problem and a highly nonlinear Continuous Stirred Tank Reactor (CSTR) process control task. 
Comparisons are conducted between the standard BLS and the proposed S-BLS under various noise conditions to demonstrate the robustness and sparsity of our method.

\subsection{Experimental Setup}
All experiments were implemented in Python on a workstation equipped with an Intel Core i7 CPU and 32GB RAM. 
To quantify the prediction performance, we utilize the Root Mean Square Error (RMSE):
\begin{equation}
RMSE = \sqrt{\frac{1}{N_{test}} \sum_{k=1}^{N_{test}} (y_k - \hat{y}_k)^2}
\end{equation}
where $y_k$ and $\hat{y}_k$ represent the ground truth and the predicted output, respectively.
Additionally, we define the \textbf{Sparsity Ratio} to measure model compactness:
\begin{equation}
\text{Sparsity} = \left( 1 - \frac{N_{active}}{N_{total}} \right) \times 100\%
\end{equation}
where $N_{active}$ is the number of non-zero weights in the output layer.

\subsection{Hyperparameter Settings}
The hyperparameters for both experiments are detailed in Table \ref{tab:hyperparams}. 
We increased the number of nodes compared to typical settings to simulate an over-parameterized environment, where standard BLS is prone to overfitting noise. 
The pruning threshold $\lambda$ (or sparsity target) for S-BLS was selected based on cross-validation on the training set.

\begin{table}[htbp]
\centering
\caption{Hyperparameter Settings for Experiments}
\label{tab:hyperparams}
\renewcommand{\arraystretch}{1.2}
\begin{tabular}{lcc}
\toprule
\textbf{Parameter} & \textbf{Nonlinear System} & \textbf{CSTR Process} \\
\midrule
Feature Nodes ($N_f \times N_g$) & $30 \times 10$ (300) & $20 \times 10$ (200) \\
Enhancement Nodes ($N_e$) & 200 & 200 \\
\textbf{Total Initial Nodes} & \textbf{501} (w/ bias) & \textbf{401} (w/ bias) \\
Regularization ($\beta$) & 0.01 & $10^{-8}$ \\
Sparsity Target (Ratio) & 0.6 (60\%) & 0.5 (50\%) \\
Training Samples & 2000 & 2000 \\
\bottomrule
\end{tabular}
\end{table}
\subsection{Case Study 1: Numerical Nonlinear System}
The first task involves identifying a nonlinear dynamic system described by the following difference equation:
\begin{equation}
y(n) = \frac{y(n-1)y(n-2)(y(n-1) + 2.5)}{1 + y^2(n-1) + y^2(n-2)} + u(n-1)
\label{eq:nonlinear_sys}
\end{equation}
where the input $u(n) \in [-2, 2]$ is uniformly distributed for training, and $u(n) = \sin(2\pi n/25)$ for testing.
To simulate harsh industrial environments, we introduce uniform noise with varying intensities $\gamma \in \{0.1, 0.2, 0.3, 0.4\}$ into the training data.

\subsubsection{Results and Analysis}
Table \ref{tab:nonlinear_results} summarizes the performance comparison. 
It is observed that the standard BLS suffers from performance fluctuations due to its tendency to fit all data points, including outliers and noise. 
In contrast, S-BLS consistently outperforms the standard BLS across all noise levels. 
Notably, as the noise level increases to 0.4, S-BLS demonstrates robust generalization capabilities, reducing the RMSE from 0.2528 (Standard BLS) to \textbf{0.1632}.

\begin{figure}[htbp]
\centering
\includegraphics[width=0.95\linewidth]{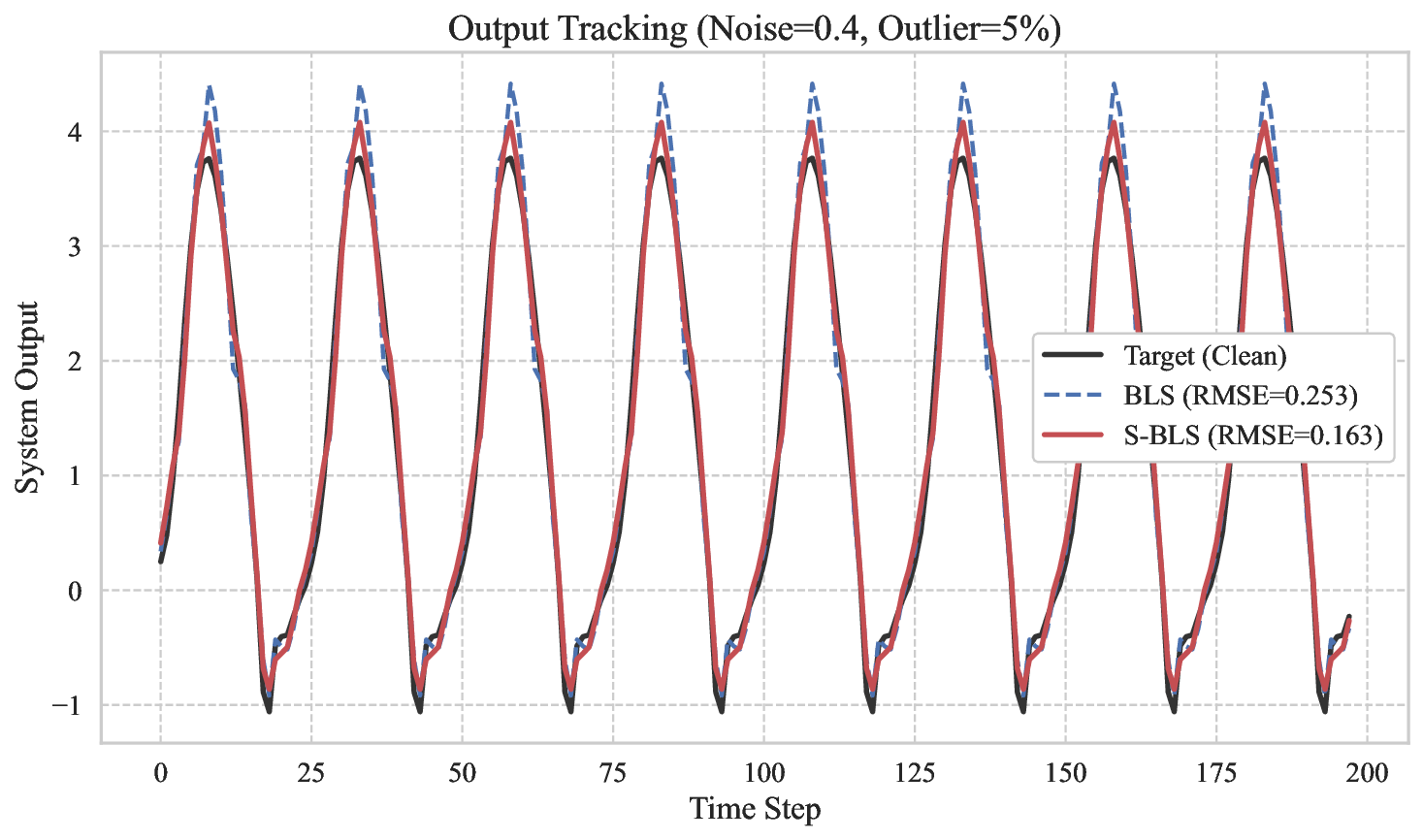}
\caption{Output tracking performance of the nonlinear system under noise level 0.4. S-BLS demonstrates superior smoothness and tracking accuracy compared to Standard BLS.}
\label{fig:tracking_03}
\end{figure}

More importantly, S-BLS achieves this accuracy by pruning approximately \textbf{50\%} of the redundant nodes, retaining only about \textbf{201 active nodes} out of 401. 
This indicates that the proposed STLS algorithm successfully identifies the compact structure of the system dynamics, effectively filtering out connections that were merely fitting the random noise.
Fig. \ref{fig:tracking_03} illustrates the output tracking performance under 0.3 noise level. The S-BLS trajectory (red) follows the ground truth (black) more closely than the standard BLS (blue), which exhibits deviations caused by noise fitting.

\begin{table}[htbp]
\centering
\caption{Performance Comparison on Nonlinear System under Varying Noise Levels}
\label{tab:nonlinear_results}
\renewcommand{\arraystretch}{1.2}
\setlength{\tabcolsep}{4pt} 
\begin{tabular}{c|cc|cc}
\toprule
\textbf{Noise} & \multicolumn{2}{c|}{\textbf{RMSE (Error)}} & \multicolumn{2}{c}{\textbf{Active Nodes (Sparsity)}} \\
\cline{2-5}
\textbf{Level} & Standard BLS & \textbf{S-BLS (Ours)} & Total & \textbf{Active} \\
\midrule
0.1 & 0.3370 & \textbf{0.2588} & 401 & 201 (49.9\%) \\
0.2 & 0.2222 & \textbf{0.1673} & 401 & 201 (49.9\%) \\
0.3 & 0.2361 & \textbf{0.1648} & 401 & 201 (49.9\%) \\
0.4 & 0.2528 & \textbf{0.1632} & 401 & 201 (49.9\%) \\
\bottomrule
\end{tabular}
\end{table}

\subsection{Case Study 2: CSTR Process Benchmark}
The Continuous Stirred Tank Reactor (CSTR) is a classic benchmark in chemical process control, characterized by strong nonlinearity and instability. 
The dynamics are governed by the mass and energy balance equations:
\begin{equation}
\begin{split}
\dot{C}_A &= \frac{q}{V}(C_{Af} - C_A) - k_0 \exp\left(-\frac{E}{RT}\right)C_A \\
\dot{T} &= \frac{q}{V}(T_f - T) + \frac{-\Delta H}{\rho C_p}k_0 \exp\left(-\frac{E}{RT}\right)C_A \\
        &\quad + \frac{UA}{V\rho C_p}(T_c - T)
\end{split}
\label{eq:cstr}
\end{equation}
where $C_A$ is the concentration of product A, $T$ is the reactor temperature, and $T_c$ is the coolant temperature (control input).
The physical parameters are defined as follows: $q$ is the feed flow rate ($100 L/min$), $V$ is the reactor volume ($100 L$), $k_0$ is the reaction rate constant ($7.2 \times 10^{10} min^{-1}$), $E$ is the activation energy ($8.75 \times 10^4 J/mol$), $R$ is the universal gas constant, $\Delta H$ is the heat of reaction, $\rho$ and $C_p$ are the density and specific heat capacity, and $UA$ is the heat transfer coefficient.
We generated 2000 training samples with additional outliers to test robustness.

\subsubsection{Results and Analysis}
The comparison results on the CSTR dataset are presented in Table \ref{tab:cstr_results}.
The proposed S-BLS consistently outperforms the standard BLS across all noise levels.
A key finding is the \textbf{high sparsity ratio}: S-BLS retains only \textbf{61 active nodes} out of 201 (pruning ratio $\approx 70\%$), yet achieves lower prediction errors (e.g., 0.0590 vs 0.0644 at noise 0.2).

This result confirms that the full-connected structure of standard BLS contains significant redundancy when modeling physical systems like CSTR. 
By forcing the weights of irrelevant nodes to zero, S-BLS not only reduces the computational burden for future inference but also effectively ignores the "noisy features" that do not contribute to the underlying physical dynamics. This makes S-BLS particularly suitable for industrial applications where interpretability and robustness are paramount.

\begin{table}[htbp]
\centering
\caption{Performance and Sparsity Comparison on CSTR Benchmark}
\label{tab:cstr_results}
\renewcommand{\arraystretch}{1.2}
\begin{tabular}{c|cc|cc}
\toprule
\textbf{Noise} & \multicolumn{2}{c|}{\textbf{RMSE (Error)}} & \multicolumn{2}{c}{\textbf{Compactness}} \\
\cline{2-5}
\textbf{Level} & Standard BLS & \textbf{S-BLS (Ours)} & Total & \textbf{Active Nodes} \\
\midrule
0.2 & 0.0644 & \textbf{0.0590} & 201 & \textbf{61} \\
0.3 & 0.0886 & \textbf{0.0809} & 201 & \textbf{61} \\
0.4 & 0.1111 & \textbf{0.1031} & 201 & \textbf{61} \\
\bottomrule
\end{tabular}
\end{table}

In summary, both experiments confirm that integrating the STLS pruning strategy into BLS significantly enhances noise immunity and model interpretability without sacrificing approximation accuracy.


\section{Conclusion}
In this paper, we presented a robust Sparse 
Broad Learning System (S-BLS). 
By integrating the Sequential Thresholded 
Least-Squares algorithm, the method enforces 
sparsity in the output weights. 
Experiments on CSTR demonstrate that S-BLS 
effectively ignores noise compared to the 
standard BLS, offering a robust solution for 
industrial system identification.

\bibliographystyle{IEEEtran}
\bibliography{refs} 

@article{chen2017bls,
  author={C. P. Chen and Z. Liu},
  title={Broad Learning System: An Effective and Efficient Incremental Learning System Without the Need for Deep Architecture},
  journal={IEEE Transactions on Neural Networks and Learning Systems},
  year={2017},
  volume={29},
  number={1},
  pages={10--24},
}

@ARTICLE{StackedBLS2021,
  author={Liu, Zhulin and Chen, C. L. Philip and Feng, Shuang and Feng, Qiying and Zhang, Tong},
  journal={IEEE Transactions on Systems, Man, and Cybernetics: Systems}, 
  title={Stacked Broad Learning System: From Incremental Flatted Structure to Deep Model}, 
  year={2021},
  volume={51},
  number={1},
  pages={209-222},
  doi={10.1109/TSMC.2020.3043147}}

@ARTICLE{FBLS1020,
  author={Feng, Shuang and Chen, C.L. Philip},
  journal={IEEE Transactions on Cybernetics}, 
  title={Fuzzy Broad Learning System: A Novel Neuro-Fuzzy Model for Regression and Classification}, 
  year={2020},
  volume={50},
  number={2},
  pages={414-424},
  doi={10.1109/TCYB.2018.2857815}}

@ARTICLE{RBLS2018,
  author={Xu, Meiling and Han, Min and Chen, C. L. Philip and Qiu, Tie},
  journal={IEEE Transactions on Cybernetics}, 
  title={Recurrent Broad Learning Systems for Time Series Prediction}, 
  year={2020},
  volume={50},
  number={4},
  pages={1405-1417},
  doi={10.1109/TCYB.2018.2863020}}

@article{weightedBLS2019,
  title={Weighted broad learning system and its application in nonlinear industrial process modeling},
  author={Chu, Fei and Liang, Tao and Chen, CL Philip and Wang, Xuesong and Ma, Xiaoping},
  journal={IEEE transactions on neural networks and learning systems},
  volume={31},
  number={8},
  pages={3017--3031},
  year={2019},
  publisher={IEEE}
}

@article{MCCBLS2020,
  title={Broad learning system based on maximum correntropy criterion},
  author={Zheng, Yunfei and Chen, Badong and Wang, Shiyuan and Wang, Weiqun},
  journal={IEEE Transactions on Neural Networks and Learning Systems},
  volume={32},
  number={7},
  pages={3083--3097},
  year={2020},
  publisher={IEEE}
}

@article{compactbls2023,
  title={Compact broad learning system based on fused lasso and smooth lasso},
  author={Chu, Fei and Liang, Tao and Chen, CL Philip and Wang, Xuesong and Ma, Xiaoping},
  journal={IEEE Transactions on Cybernetics},
  volume={54},
  number={1},
  pages={435--448},
  year={2023},
  publisher={IEEE}
}

@article{sindy2016,
  title={Discovering governing equations from data by sparse identification of nonlinear dynamical systems},
  author={Brunton, Steven L and Proctor, Joshua L and Kutz, J Nathan},
  journal={Proceedings of the national academy of sciences},
  volume={113},
  number={15},
  pages={3932--3937},
  year={2016},
  publisher={National Academy of Sciences}
}

@article{RVFLNN1994,
  title={Learning and generalization characteristics of the random vector functional-link net},
  author={Pao, Yoh-Han and Park, Gwang-Hoon and Sobajic, Dejan J},
  journal={Neurocomputing},
  volume={6},
  number={2},
  pages={163--180},
  year={1994},
  publisher={Elsevier}
}

@article{CSBLS2023,
title = {Learning broad learning system with controllable sparsity through L0 regularization},
journal = {Applied Soft Computing},
volume = {136},
pages = {110068},
year = {2023},
issn = {1568-4946},
doi = {https://doi.org/10.1016/j.asoc.2023.110068},
url = {https://www.sciencedirect.com/science/article/pii/S1568494623000868},
author = {Fei Chu and Guanghui Wang and Jun Wang and C.L. Philip Chen and Xuesong Wang}
}

\end{document}